# Evolution of the genetic code. Emergence of DNA

Semenov D.A. ([dasem@mail.ru](dasem@mail.ru))

International Research Center for Studies of Extreme States of the Organism at the Presidium of the Krasnoyarsk Research Center, Siberian Branch of the Russian Academy of Sciences

*Abstract. This hypothesis can provide an opportunity to trace logically the process of the emergence of the DNA double helix. AT-enrichment in this hypothesis is main factor of evolution of DNA double helix from RNA double helix.*

The hypothesis that at the early stages there were just cytosine and guanine [1,2,3] makes the picture of the RNA world much poorer. It is difficult to form a great variety of tertiary structures based on nucleotides that tend to form almost exclusively Watson-Crick's pairs. This hypothesis can be said to restrict and, thus, to simplify the picture of the early RNA world.

However, this hypothesis can provide an opportunity to trace logically the process of the emergence of the DNA double helix. The generally accepted picture of the RNA world suggests that DNA was invented to separate the function of information storage from the function of translation. The double copy is more stable towards interferences and mutations, but less accessible to enzymes. The emergence of DNA is often associated with the genome increase, suggesting that smaller organisms such as viruses can make do with RNA. This interpretation ignores one low-probability assumption – it is accepted de fide that a ***random*** replacement of ribose by deoxyribose resulted in the transformation of the previously single-strand RNA into the double helix. To make the copy more stable towards mutations! But this is pure teleology!

Yet, nucleotides themselves already possess complementarity properties that allow them (in a miraculous way) to be incorporated into this double helix. If this did not occur at the early stages and complementarity was necessary for the nucleotide-to-nucleotide replication, how can one account for this amazing correspondence?

The hypothesis suggesting that at early stages of the evolution of the genetic code there were just cytosine and guanine makes it possible to bring the emergence of DNA back to evolutionary philosophy.

The complementary long chains based on guanine and cytosine only (CG-polyribonucleotides) are more apt to form the double helix than the complementary polyribonucleotides based on all four pairs. The reason is simple: guanine and cytosine form just Watson-Crick's pairs due to a significant energy benefit as compared to other variants.

I can admit that different variants of oligomeric complementary RNAs based on cytosine and guanine can yield double helixes of different stabilities. It is important that for a random sequence of CG ribo-oligomer the RNA double helix will be stable.

The next step in my investigation will be emergence of uracil due to cytosine deamination. Deamination first leads to the formation of uracil in the enol form. Then it must be transformed into the keto form, which is more stable in solution. However, when uracil is gripped in the backbone of the RNA double helix, the enol form is more advantageous for it. Uracil in the enol form can make two and even three hydrogen bonds with guanine, while uracil in the keto form would be able to make only one hydrogen bond. One can say that the enol form of uracil is stabilized by its interaction with the complementary guanine. The keto-enol tautomerism is a well-studied process and chemists will understand me without any further proofs. I should add that Watson first considered nucleotide formulae in the enol form as the more probable, guided by reference books of that time [4]. Fortunately, I can give convincing arguments using commonly available information. Codon-anticodon interaction results in the formation of a short segment of the double helix. In the case of GC-rich codons, opposite to uracil in the third position of the codon there is inosine in the anticodone. Crick's wobble hypothesis [5] allows a solution for this pair only by wobbling the third nucleotides of the codon and the anticodon. However, in the double helix, their position is also stabilized by stacking. If stacking exerts significant influence so that wobbling becomes impossible, then no uracil-inosine complementarity is possible for the keto form of uracil, i.e. there cannot be any hydrogen bonds. In this case, uracil is in the state that most closely imitates cytosine.

Although Crick's idea is certainly true and proved to be fruitful, it is superfluous for explaining the codon multivariant pairing. If C-I are Watson-Crick's pairs and U-I pairs are formed in accordance with the wobble hypothesis, it is not clear why these codons are always indistinguishable, although they have different conformations, which can be stabilized. With the keto-enol tautomerism, both codons have the same conformation.

Why is uracil opposed by inosine is some anticodons and by guanine in others? According to Crick's hypothesis inosine and guanine are indistinguishable to uracil. If we suppose the presence of the enol form of uracil, we can suggest that inosine emerged in anticodons in an evolutionary way because in this case the guanine amino group could not form a hydrogen bond. That is, in this case inosine (guanine without an amino group) is sufficient. I should add that the enol form of uracil can be registered in NMR spectra, which makes the necessary experiments easy to perform.

Uracil incorporated into the RNA double helix is very similar to cytosine. Accumulation of the sufficient amount of uracil gradually makes the double helix unstable as the bond is not so strong. The loss of stability at a certain section is accompanied by the transformation of uracil into the keto form and breakdown of the double helix. It is the occurrence of this instability (catastrophe) that causes the system to become more complex. For uracil to continue accumulating in the double helix, the RNA double helix must grow more stable. The greater stability is attained with the "invention" of adenine, which can form two hydrogen bonds with uracil in the keto form.

The emergence of adenine is a source of strain in the ribophosphate backbone of the RNA double helix. In AT-rich codons, uracil in the third position of the codon is opposed by guanine in the third position of the anticodon, suggesting a distortion in the codon conformation.

Accumulation of AU pairs leads to the already familiar catastrophe – loss of double helix stability. The way out of this catastrophe cannot be achieved by conventional means but only by the system becoming more complicated. The replacement of ribose by deoxyribose reduces the rigidity of the sugar-phosphate backbone. Thus, DNA emerges.

Cytosine methylation and emergence of thymine must have been associated with the stage at which such a catastrophe occurred. Methylation, e.g., could favor further stabilization of uracil in the keto form. Oxygen is an electron acceptor while the methyl group an electron donor, so the presence of the methyl group reduces the probability of the proton being near oxygen.

**Acknowledgement**

The author would like to thank Krasova E. for her assistance in preparing this manuscript.